\documentclass[journal,10pt]{IEEEtran}
\usepackage{amsfonts}
\usepackage{amssymb}
\usepackage{eurosym}
\usepackage{cite}
\usepackage{graphicx}
\usepackage{epstopdf}
\usepackage{amsmath}
\usepackage{tikz,lipsum}
\usepackage{caption}
\usepackage[T1]{fontenc}
\usepackage{amsthm}
\usepackage{mathrsfs}
\usepackage{slashbox}
\usepackage{mathrsfs}
\usepackage{color}
\usepackage{stfloats}
\usepackage{xcolor, etoolbox}
\usepackage[papersize={8.5in,11in}]{geometry}
\usepackage{slashbox}

\makeatletter
\pretocmd\@bibitem{\color{black}\csname keycolor#1\endcsname}{}{\fail}
\newcommand\citecolor[1]{\@namedef{keycolor#1}{\color{black}}}
\makeatother
\citecolor{w1}
\citecolor{w2}
\IEEEoverridecommandlockouts
\newcounter{mytempeqncnt}
\newtheorem{remark}{Remark}

\newtheorem{proposition}{Proposition}
\newtheorem{corollary}{Corollary}

\begin{document}

\title{On the Distribution of the Sum of M\'{a}laga-$\mathcal{M}$ Random
Variables and Applications}
\author{Elmehdi Illi, \IEEEmembership{Member, IEEE}, Faissal El Bouanani, %
\IEEEmembership{Senior Member, IEEE}, and Fouad Ayoub,
\IEEEmembership{Member,
IEEE} \thanks{%
E. Illi and F. El Bouanani are with ENSIAS College of Engineering, Mohammed
V University, Rabat, Morocco (e-mails: \{elmehdi.illi,
f.elbouanani\}@um5s.net.ma).} \thanks{%
F. Ayoub is with CRMEF, Kenitra, Morocco (e-mail: ayoub@crmefk.ma).}}
\maketitle

\begin{abstract}
In this {{paper}}, a very accurate approximation method for the
statistics of the sum of M\'{a}laga-$\mathcal{M}$ random variates with
pointing error (MRVs) is proposed. In particular, the probability density
function\ of MRV is approximated by a Fox's $H$-function through the
moment-based approach. Then, the respective moment-generating function of
the sum of $N$ MRVs is provided, based on which the average symbol error
rate is evaluated for an $N $-branch maximal-ratio combining (MRC) receiver.
The retrieved results show that the proposed approximate results match
accurately with the exact simulated ones. Additionally, the results show
that the achievable diversity order increases as a function of the number of
MRC diversity branches.
\end{abstract}



\IEEEpeerreviewmaketitle

\section{Introduction}

In recent years, there has been an increasing interest in deriving the
statistical properties of the sum of random variates (RVs), namely the
probability density function (PDF) and the moment-generating function (MGF).
Such statistical properties are of paramount importance in analyzing the
performance of wireless communication systems (WCSs) employing
multiple-input multiple-output (MIMO) schemes and maximal-ratio combining
(MRC) diversity technique.

M\'{a}laga-$\mathcal{M}$ distribution has been widely advocated as a
universal model for representing the atmospheric turbulence impairment in
free-space optical (FSO) links \cite{ansarimalaga, paris}, as it generalizes
several turbulence-induced fading models (e.g., {{shadowed-Rician}%
}, Gamma-Gamma, Log-Normal, double Weibull) {{\cite{paris}}}.
From another front, there has been a rising attention in the analysis of
MIMO\ FSO\ WCSs employing MRC combining scheme, where the output SNR\ is the
sum of the SNRs on the receiver branches. For instance, the authors in \cite%
{ruiz} dealt with the capacity performance of a multiple-input single-output
(MISO) FSO system employing equal-gain combining technique, subject to
Gamma-Gamma fading and {{pointing error impairment ({{PEI}})}}. Additionally, the
authors in \cite{ma} dealt with the outage and average bit error rate
performance for both MRC and selection combining techniques over Gamma-Gamma
fading. Importantly, the work in \cite{tansal} analyzed the performance of
an $M$-branch MISO FSO system using maximal-ratio transmission, where the
FSO links\ undergo M\'{a}laga-$\mathcal{M}$ fading without {{PEI}}. {{On the other hand,
other works such as \cite{w1,w2} dealt with the distribution of the product
of shadowed-Rician RVs.}}

{{From the above-mentioned works, the authors assessed the performance of FSO\
systems employing either MRC or EGC receivers, subject to either Gamma-Gamma with {{PEI}}
or M\'{a}laga-$\mathcal{M}$ fading model without {{PEI}}.
To the best of the authors' knowledge, neither the sum of M\'{a}laga-$\mathcal{M}$ RVs with {{PEI}} (MRVs) nor the analytical performance
of FSO system employing MRC\ receiver, subject to M\'{a}laga-$\mathcal{M}$ fading
with {{PEI}} have been investigated before in the literature.}}
Throughout this {{paper}}, and distinctly from the works \cite%
{tansal, ruiz, newarticle}, we aim at proposing a highly-accurate
approximation {{for the statistics of}} the sum of MRVs. In
particular, we approximate the PDF of MRV through a Meijer's $G$-function,
by the use of the moments-based approach. Distinct from \cite{wcl}, the
first six moments of the distribution are evaluated instead of only the
first five ones, so as to enhance the approximation accuracy. Capitalizing
on this result, the MGF of the sum of MRVs is retrieved, based on which the
respective average symbol error rate (ASER) for a single-input
multiple-output (SIMO) FSO system employing MRC scheme is evaluated in
approximate and asymptotic forms. The derived MGF and ASER results are
provided for two cases: (i) independent and identically-distributed (i.i.d),
and (ii) independent and non-identically distributed (i.n.i.d) MRVs. {%
{Importantly, the derived results are generalizing the ones of an FSO system
employing MRC receiver, subject either to traditional M\'{a}laga-$\mathcal{M}
$ fading model without {{PEI}}, investigated in \cite{tansal}, or
subject to Gamma-Gamma fading model with {{PEI}}, inspected in \cite{ruiz, newarticle}.}}

The main contributions of this work are given as follows: First, we propose
an accurate approximate expression of the MRV's PDF. Then, we retrieve the
MGF of the sum of MRVs for both i.i.d and i.n.i.d cases. Based on these
last-mentioned results, we derive the ASER and its asymptotic expressions
for various modulations of an $N$-branch MRC receiver in approximate form,
for both aforementioned cases. It is further demonstrated that the
achievable diversity order is increasing with the increase in the number of
MRC branches.

\section{Proposed Approximation}

In this section, a simple and accurate approximate PDF for the MRV is
presented, based on which the MGF of the sum of MRVs is derived, for both
i.i.d and i.n.i.d cases.

\subsection{Probability Density Function}

Let $\gamma $ be an MRV encompassing the atmospheric turbulence and the
{{PEI}} with PDF \cite{ansarimalaga}\footnote{%
We consider in this work the M\'{a}laga-$\mathcal{M}$ distribution as in
\cite{ansarimalaga} for the case of coherent heterodyne detection (i.e., $%
r=1 $).}
\begin{equation}
f_{\gamma }\left( x\right) =\frac{\xi ^{2}A}{2x}\sum_{m=1}^{\beta
}b_{m}G_{1,3}^{3,0}\left( \frac{Bx}{\mu _{1}}\left\vert
\begin{array}{c}
-;\xi ^{2}+1 \\
\xi ^{2},\alpha ,m;-%
\end{array}%
\right. \right) ,x>0;  \label{malagapdf}
\end{equation}%
where $A=\frac{2\alpha {}^{\frac{\alpha }{2}}}{h_{j}^{1+\frac{\alpha }{2}%
}\Gamma (\alpha )}\left( \frac{\beta }{\beta +\frac{\Omega ^{\prime }}{h}}%
\right) ^{\beta +\frac{\alpha }{2}},$ $\alpha $ and $\beta $ are the
atmospheric turbulence severity parameters, $h=2d_{o}\left( 1-\epsilon
\right) $ and $\Omega ^{^{\prime }}=\Omega +2d_{0}\epsilon +2\sqrt{%
2d_{0}\epsilon \Omega }\cos \left( \Theta _{A}-\Theta _{B}\right) $ denote
the average power of the scattering component received by off-axis eddies
and the one of the coherent contributions, respectively, with $2d_{0}$ is
the average power of the total scatter components, $\epsilon $ is the amount
of scattering power coupled-to-LOS\ component, $\Omega $ is the average
power of the dominant LOS\ component, and $\Theta _{A}$ and $\Theta _{B}$
are deterministic phases of the LOS and coupled-to-LOS terms, respectively.
Besides, $G_{p,q}^{m,n}\left( .\left\vert .\right. \right) $ refers to the
Meijer's $G$-function \cite{kilbas}, $B=\frac{\xi ^{2}\alpha \beta \left(
h+\Omega ^{^{\prime }}\right) }{\left( \xi ^{2}+1\right) \left( h\beta
+\Omega ^{^{\prime }}\right) },$ $\xi $ denotes the {{PEI}} severity
parameter, and $\mu _{1}=\mathbb{E}\left[ \gamma \right] $ stands for the
average value of $\gamma $, where $\mathbb{E}\left[ .\right] $ refers to the
expectation operator \cite{illi}. Furthermore, $b_{m}=\binom{\beta -1}{m-1}%
\frac{\left( h\beta +\Omega ^{\prime }\right) ^{1+\frac{\alpha }{2}}}{(m-1)!}%
\left( \frac{\Omega ^{\prime }}{h}\right) ^{m-1}\beta ^{-\frac{\alpha }{2}%
-m}\alpha ^{-\frac{\alpha }{2}}.$

\begin{proposition}
The PDF\ of $\gamma $ can be approximated accurately as follows%
\begin{equation}
f_{\gamma }\left( x\right) \approx a_{1}G_{2,2}^{2,0}\left( \frac{x}{a_{2}}%
\left\vert
\begin{array}{c}
-;a_{3},a_{4} \\
a_{5},a_{6};-%
\end{array}%
\right. \right) ,  \label{pdfapprox}
\end{equation}%
where%
\begin{equation}
a_{1}=\frac{\Gamma \left( a_{3}+1\right) \Gamma \left( a_{4}+1\right) }{%
a_{2}\Gamma \left( a_{5}+1\right) \Gamma \left( a_{6}+1\right) },  \label{a1}
\end{equation}%
\begin{equation}
a_{2}=\frac{\mathcal{L}_{4}}{2}-\mathcal{L}_{3}+\frac{\mathcal{L}_{2}}{2},
\label{a2}
\end{equation}%
\begin{equation}
a_{3}=\frac{-4\mathcal{G}_{4}+9\mathcal{G}_{3}-6\mathcal{G}_{2}+\mathcal{G}%
_{1}}{\mathcal{G}_{4}-3\mathcal{G}_{3}+3\mathcal{G}_{2}-\mathcal{G}_{1}},
\label{a3}
\end{equation}%
\begin{equation}
a_{4}=\frac{-\phi -\sqrt{\phi ^{2}-4\left( \delta p+\lambda r\right) \left(
\lambda q+\sigma s\right) }}{2\left( \delta p+\lambda r\right) },  \label{a4}
\end{equation}%
\begin{equation}
a_{5}=\frac{\kappa -\sqrt{\kappa ^{2}-4\eta }}{2}-1,  \label{a5}
\end{equation}%
\begin{equation}
a_{6}=\frac{\kappa +\sqrt{\kappa ^{2}-4\eta }}{2}-1,  \label{a6}
\end{equation}%
with%
\begin{equation}
\mathcal{L}_{i}=\varphi _{i}\left( a_{4}+i\right) \left( a_{3}+i\right) ,
\label{li}
\end{equation}%
\begin{equation}
\mathcal{G}_{i}=\varphi _{i}\left( a_{4}+i\right) ,  \label{gi}
\end{equation}%
\begin{equation}
\phi =\lambda \left( p+s\right) +\sigma r+\delta q,  \label{phi}
\end{equation}%
\begin{equation}
\lambda =5\varphi _{5}-12\varphi _{4}+9\varphi _{3}-2\varphi _{2},
\label{landa}
\end{equation}%
\begin{equation}
p=-4\varphi _{4}+9\varphi _{3}-6\varphi _{2}+\mu _{1},  \label{pp}
\end{equation}%
\begin{equation}
s=4\varphi _{4}-9\varphi _{3}+6\varphi _{2}-\mu _{1},  \label{ss}
\end{equation}%
\begin{equation}
\sigma =25\varphi _{5}-48\varphi _{4}+27\varphi _{3}-4\varphi _{2},
\end{equation}%
\begin{equation}
r=\varphi _{4}-3\varphi _{3}+3\varphi _{2}-\mu _{1},
\end{equation}%
\begin{equation}
\delta =\varphi _{5}-3\varphi _{4}+3\varphi _{3}-\varphi _{2},  \label{delta}
\end{equation}%
\begin{equation}
q=\mu _{1}-16\varphi _{4}+27\varphi _{3}-12\varphi _{2},  \label{qq}
\end{equation}%
\begin{equation}
\kappa =\frac{\mathcal{L}_{2}-\mathcal{L}_{1}}{a_{2}}-1,  \label{kappa}
\end{equation}%
\begin{equation}
\eta =\frac{\mathcal{L}_{1}}{a_{2}},  \label{eta}
\end{equation}%
and $\varphi _{i}=\frac{\mu _{i}}{\mu _{i-1}}\left( i\geq 1\right) ,$ $\mu
_{i}$ is the $i$-th moment of $\gamma $, and $\Gamma \left( .\right) $ is
the Gamma function \cite[Eq. (8.350.1)]{integrals}.

\begin{remark}
Interestingly, such approximation with all the above parameters remains
accurate for other kinds of distributions.
\end{remark}

{%
{\begin{IEEEproof}
The proof is provided in Appendix A.
\end{IEEEproof}}}
\end{proposition}

\begin{remark}
\label{proportion} First, it is clearly seen that $\mu _{i}$ is proportional
to $\mu _{1}^{i}.$ It follows that $\varphi _{i},$ and consequently the
parameters defined in (\ref{landa})-(\ref{delta}), are proportional to $\mu
_{1},$ while $\phi $ is proportional to $\mu _{1}^{2}.$ Thus, $a_{4},$ given
in (\ref{a4}), is unitless. Therefore, it yields from (\ref{a3}) and (\ref%
{gi}) that $a_{3}$ is also unitless. Furthermore, one can notice also from (%
\ref{a2}) and (\ref{li}) that $a_{2}$ is proportional to $\mu _{1},$ and
that $\kappa $ and $\eta $ are unitless. Lastly, $a_{5}$ and $a_{6}$ are
then unitless, and $a_{1}$ is inversely proportional to $\mu _{1}$.
\end{remark}

\subsection{Moment-Generating Function}

\begin{corollary}
The MGF\ of\ $\gamma $ can be approximated as follows%
\begin{equation}
M_{\gamma }(s)\approx \frac{a_{1}}{s}G_{3,2}^{2,1}\left( \frac{1}{sa_{2}}%
\left\vert
\begin{array}{c}
0;a_{3},a_{4} \\
a_{5},a_{6};-%
\end{array}%
\right. \right) ;s>0.  \label{mgfapprox}
\end{equation}

\begin{IEEEproof}
The MGF\ of $\gamma $ can be evaluated as follows%
\begin{equation}
M_{\gamma }(s)=\int_{0}^{\infty }e^{-sx}f_{\gamma }(x)dx.  \label{mgfdef}
\end{equation}

By plugging (\ref{pdfapprox}) into (\ref{mgfdef}), and making use of the
Laplace transform \cite[Eq. (2.19)]{mathai}, (\ref{mgfapprox}) is reached.
\end{IEEEproof}
\end{corollary}

\begin{proposition}
Let us consider the set $\left\{ \gamma _{j}\right\} _{1\leq j\leq N}$ of
MRVs with parameters $\alpha ^{(j)},$ $\beta ^{(j)}$, $\xi ^{(j)},$ and $\mu
_{1}^{(j)}$. The MGF\ of\ $\gamma _{T}=\sum\limits_{j=1}^{N}\gamma _{j}$ can
be expressed in an approximate form as%
\begin{align}
M_{\gamma _{T}}^{(\text{i.i.d})}(s)& \approx N!\left( \frac{a_{1}}{s}\right)
^{N}\sum_{k_{1}+k_{2}=N}\frac{\left( k_{1}!k_{2}!\right) ^{-1}}{%
a_{2}^{\varkappa _{k_{1},k_{2}}}}  \notag \\
& \times \sum\limits_{l=0}^{\infty
}\sum\limits_{q_{1}+q_{2}=l}c_{q_{1}}^{(1)}c_{q_{2}}^{(2)}s^{-\varrho
_{l,k_{1},k_{2}}},  \label{mgftotal}
\end{align}%
and
\begin{align}
M_{\gamma _{T}}^{(\text{i.n.i.d})}(s)& \approx \sum_{\substack{ %
k_{1}+k_{2}+...+k_{N}\geq N  \\ k_{j}=1,2}}\prod\limits_{j=1}^{N}\left(
a_{1,j}a_{2,j}^{-a_{k_{j}+4,j}}\right)  \notag \\
& \times \sum\limits_{l=0}^{\infty }s^{-\upsilon
-N-l}\sum\limits_{q_{1}+q_{2}+...+q_{N}=l}\prod%
\limits_{i=1}^{N}b_{q_{i},i}^{(k_{i})},  \label{mgftotalinid}
\end{align}%
for i.i.d and i.n.i.d cases, respectively, where $\varkappa
_{k_{1},k_{2}}=a_{5}k_{1}+a_{6}k_{2}$, $\varrho _{l,k_{1},k_{2}}=\varkappa
_{k_{1},k_{2}}+l,$ $\upsilon =\sum\limits_{j=1}^{N}a_{k_{j}+4,j},$
\begin{equation}
c_{m}^{(i)}=\left\{
\begin{array}{l}
\left( b_{0}^{(i)}\right) ^{k_{i}},m=0 \\
\frac{1}{mb_{0}^{(i)}}\sum\limits_{l=1}^{m}\left( lk_{i}-m+l\right)
b_{l}^{(i)}c_{m-l}^{(i)};m\geq 1%
\end{array}%
\right. ,  \label{ci}
\end{equation}%
and%
\begin{equation}
b_{l}^{(i)}=\frac{(-1)^{l}\Gamma \left( 1+a_{i+4}+l\right) \Gamma \left(
\left( 3-2i\right) \left( a_{6}-a_{5}\right) -l\right) }{l!\Gamma \left(
a_{3}-a_{i+4}-l\right) \Gamma \left( a_{4}-a_{i+4}-l\right) a_{2}^{l}}.
\label{b1}
\end{equation}

\begin{remark}
\label{bl}$a_{i,j}$ and $b_{l,j}^{\mathbf{(}k\mathbf{)}}$ represent the
coefficients $a_{i}$ and $b_{l}^{\mathbf{(}k\mathbf{)}}$ for the i.n.i.d
case, respectively with $j=1,..,N.$
\end{remark}

\begin{IEEEproof}
{%
{The MGF in (\ref{mgfapprox}) can be written through the Mellin-Barnes
definition as \cite[Eq. (1.112)]{mathai}{\small\begin{equation}
M_{\gamma _{j}}(s)\approx \frac{a_{1,j}}{2\pi is}\int_{C_{t}}\frac{\Gamma
\left( a_{5,j}+t\right) \Gamma \left( a_{6,j}+t\right) \Gamma \left(
1-t\right) }{\Gamma \left( a_{3,j}+t\right) \Gamma \left( a_{4,j}+t\right) \left(sa_{2,j}\right)^{-t} }dt,  \label{mgffff}
\end{equation}}where $i^{2}=-1.$ Using the residues theorem \cite[Theorem 1.2]{kilbas}, the
MGF in (\ref{mgffff}) can be written as the summation of the residues
evaluated at the left poles of the associated integrand function as \cite[Eqs. (1.3.5), (1.3.6)]{kilbas}\begin{equation}
M_{\gamma _{j}}(s)=\frac{a_{1,j}\left( \Delta _{1,j}+\Delta _{2,j}\right) }{s},  \label{mgfapproxr}
\end{equation}with
\begin{equation}
\Delta _{k,j}=\left( sa_{2,j}\right) ^{-a_{k+4,j}}\sum\limits_{l=0}^{\infty
}b_{l,j}^{(k)}s^{-l};k=1,2,  \label{delta1}
\end{equation}}}and $b_{l,j}^{(k)}$ being defined similarly to $%
b_{l}^{(k)} $ in (\ref{bl}) by replacing $a_{i}$ by $a_{i,j}$. On the other
hand, since $\gamma _{T}=\sum\limits_{j=1}^{N}\gamma _{j},$ the MGF\ of $%
\gamma _{T}$ is expressed as%
\begin{eqnarray}
M_{\gamma _{T}}^{\text{(i.i.d)}}(s) &=&\left[ M_{\gamma _{j}}(s)\right] ^{N},
\label{mgfrelation} \\
M_{\gamma _{T}}^{\text{(i.n.i.d)}}(s) &=&\prod\limits_{j=1}^{N}M_{\gamma
_{j}}(s),  \label{mgfrelation2}
\end{eqnarray}%
for the i.i.d and i.n.i.d cases, respectively. Furthermore, by assuming $%
a_{i,j}=a_{i}$ and $b_{l,j}^{(k)}=b_{l}^{(k)}$ for the i.i.d case, and using
the multinomial theorem as well as \cite[Eq. (0.314)]{integrals} alongside
with some algebraic manipulations, (\ref{mgftotal}) is attained. Finally, by
rearranging the products of $\Delta _{k,j}$ $(k=1,2,j=1,..,N)$ and
retrieving the scale factors of $s^{-l}$, (\ref{mgftotalinid}) is obtained.
\end{IEEEproof}
\end{proposition}

\section{Error Rate Analysis}

{%
{In this section, and capitalizing on the previously derived
results, the ASER\ of a SIMO\ FSO-based WCS, employing MRC\ technique, is
inspected. In the considered system, an optical transmitter, consisting of a
LED/Laser, transmits an optical beam through a turbulent channel toward a
receiver equipped by several optical apertures. In particular, the ASER\ of
the considered WCS, for various coherent modulations is inspected under both
i.i.d and i.n.i.d cases. Also, asymptotic representations of this metric
over both cases are derived.}}

\subsection{Approximate Analysis}

\begin{proposition}
The ASER\ for various modulations and $N$-branch MRC\ receiver, subject to M%
\'{a}laga-$\mathcal{M}$ turbulence-induced fading with {{PEI}}, can
be approximated as%
\begin{align}
\overline{P}_{s}^{(\text{i.i.d})}& \approx \frac{\rho }{\pi }\left( \frac{%
4a_{1}}{\theta }\right) ^{N}\sum_{k_{1}+k_{2}=N}\frac{N!\left( \theta
a_{2}\right) ^{-\varkappa _{k_{1},k_{2}}}}{k_{1}!k_{2}!}  \notag \\
& \times \sum\limits_{l=0}^{\infty
}\sum\limits_{q_{1}+q_{2}=l}c_{q_{1}}^{(1)}c_{q_{2}}^{(2)}\theta
^{-l}4^{\varrho _{l,k_{1},k_{2}}}  \notag \\
& \times \mathcal{B}\left( N+\varrho _{l,k_{1},k_{2}}+\frac{1}{2},N+\varrho
_{l,k_{1},k_{2}}+\frac{1}{2}\right) ,  \label{aserfinal}
\end{align}%
and%
\begin{align}
\overline{P}_{s}^{(\text{i.n.i.d})}& \approx \frac{\rho }{\pi }\sum
_{\substack{ k_{1}+k_{2}+...+k_{N}\geq N  \\ k_{j}=1,2}}\prod%
\limits_{j=1}^{N}\left( a_{1,j}a_{2,j}^{-a_{k_{j}+4,j}}\right)  \notag \\
& \times \sum\limits_{l=0}^{\infty }\mathcal{B}\left( \upsilon +N+l+\frac{1}{%
2},\upsilon +N+l+\frac{1}{2}\right)  \notag \\
& \times \left( \frac{4}{\theta }\right) ^{\upsilon
+N+l}\sum\limits_{q_{1}+q_{2}+...+q_{N}=l}\prod%
\limits_{j=1}^{N}b_{q_{j},j}^{(k_{j})},  \label{aserfinalinid}
\end{align}%
for the i.i.d and i.n.i.d cases, respectively, where $\rho $ and $\theta $
are modulation-dependent parameters, and $\mathcal{B}\left( .,.\right) $
refers to the beta function \cite[Eq. (8.384.1)]{integrals}.

\begin{IEEEproof}
{%
{The ASER\ can be expressed using the MGF as \cite[Eq. (23)]{illi}\begin{equation}
\overline{P}_{s}^{(\text{x})}=\frac{2\rho }{\pi }\int_{0}^{\frac{\pi }{2}}M_{\gamma _{T}}^{\text{(x)}}\left( \frac{\theta }{\sin ^{2}\phi }\right)
d\phi ,  \label{stepint}
\end{equation}with x$\in \{$i.i.d, i.n.i.d$\}$. By plugging (\ref{mgftotal})-(\ref{mgftotalinid}) into (\ref{stepint}), it yields\begin{align}
\overline{P}_{s}^{(\text{i.i.d})}& \approx \frac{2\rho }{\pi }N!\left( \frac{a_{1}}{s}\right) ^{N}\sum_{k_{1}+k_{2}=N}\frac{\left( k_{1}!k_{2}!\right)
^{-1}}{a_{2}^{\varkappa _{k_{1},k_{2}}}}  \notag \\
& \times \sum\limits_{l=0}^{\infty }\sum\limits_{q_{1}+q_{2}=l}\frac{c_{q_{1}}^{(1)}c_{q_{2}}^{(2)}}{\theta ^{\varrho _{l,k_{1},k_{2}}}}\int_{0}^{\frac{\pi }{2}}\left( \sin \phi \right) ^{2\varrho _{l,k_{1},k_{2}}}d\phi ,
\end{align}and
\begin{align}
\overline{P}_{s}^{(\text{i.n.i.d})}& \approx \frac{2\rho }{\pi }\sum
_{\substack{ k_{1}+k_{2}+...+k_{N}\geq N \\ k_{j}=1,2}}\prod\limits_{j=1}^{N}\left( a_{1,j}a_{2,j}^{-a_{k_{j}+4,j}}\right)
\sum\limits_{l=0}^{\infty }\theta ^{-\upsilon -N-l}  \notag \\
& \times \int_{0}^{\frac{\pi }{2}}\left( \sin ^{2}\phi \right) ^{2\left(
\upsilon +N+l\right) }d\phi
\sum\limits_{q_{1}+q_{2}+...+q_{N}=l}\prod\limits_{i=1}^{N}b_{q_{i},i}^{(k_{i})}.
\end{align}

Lastly, by using \cite[Eq. (3.621.1)]{integrals}, (\ref{aserfinal})-(\ref{aserfinalinid}) are reached.}%
}
\end{IEEEproof}
\end{proposition}

{%
{
\begin{remark}
For M\'alaga-$\mathcal{M}$ distribution with {{PEI}},
$\mu_1$ or $\mu^{(j)}_1$ represents the average electrical SNR per receiver's
branch. Importantly, from \textbf{Remark 2}, the higher $\mu_1$ and
$\mu^{(j)}_1$ are, the lower are $a_1$ and $a_{1,j}$ and the greater are $a_2$ and $a_{2,j}$, respectively.
Therefore, it can be seen that the MGF and ASER of the considered system, given in
(\ref{mgftotal}), (\ref{mgftotalinid}), (\ref{aserfinal}), and (\ref{aserfinalinid}) are decreasing with respect to
$\mu_1$ and $\mu^{(j)}_1$.
\end{remark}}}

\subsection{Asymptotic Analysis}

\begin{corollary}
In the high SNR\ regime (i.e., $\mu _{1}\rightarrow \infty $), the ASER\ can
be asymptotically approximated, for both i.i.d and i.n.i.d cases, as follows%
\begin{equation}
\overline{P}_{s}^{(\text{i.i.d,}\infty )}\sim G_{c}^{\text{(i.i.d)}%
}a_{2}^{-G_{d}^{\text{(i.i.d)}}},  \label{aserasm}
\end{equation}%
\begin{align}
\overline{P}_{s}^{(\text{i.n.i.d,}\infty )}& \sim \frac{\rho }{\pi }\mathcal{%
B}\left( \zeta +N+\frac{1}{2},\zeta +N+\frac{1}{2}\right)  \notag \\
& \times \left( \frac{4}{\theta }\right) ^{\zeta
+N}\prod\limits_{j=1}^{N}\left( b_{0,j}^{(g_{j})}\tau
^{(j)}a_{2,j}^{-a_{g_{j}+4,j}-1}\right) ,  \label{aserasminid}
\end{align}%
where the coding gain and achievable diversity order for the i.i.d case are
\begin{align}
G_{c}^{\text{(i.i.d)}}& =\frac{\rho }{\pi }\left( \left( \frac{4}{\theta }%
\right) ^{a_{m+4}+1}\tau b_{0}^{\mathbf{(}m\mathbf{)}}\right) ^{N}  \notag \\
& \times \mathcal{B}\left( N\left( a_{m+4}+1\right) +\frac{1}{2},N\left(
a_{m+4}+1\right) +\frac{1}{2}\right) ,  \label{Gc}
\end{align}%
and
\begin{equation}
G_{d}^{\text{(i.i.d)}}=N\left( a_{m+4}+1\right) ,  \label{gd}
\end{equation}%
$\text{respectively, with }$$\tau $$=$$\frac{\Gamma \left( a_{3}+1\right)
\Gamma \left( a_{4}+1\right) }{\Gamma \left( a_{5}+1\right) \Gamma \left(
a_{6}+1\right) }$ \footnote{$\tau $ is denoted $\tau ^{(j)}$ for the i.n.i.d
case.}, $m$=$\left\{
\begin{array}{l}
1,a_{5}<a_{6} \\
2,\text{ }a_{5}>a_{6}%
\end{array}%
\right. $ , $\zeta $$=$$\sum\limits_{j=1}^{N}a_{g_{j}+4,j}$$\text{, and }$$%
a_{g_{j}+4,j}=\min \left( a_{5,j},a_{6,j}\right) $.

\begin{IEEEproof}
When $\mu _{1}^{(j)}\rightarrow \infty $, it yields from Remark \ref%
{proportion} that $a_{1,j}$ and $a_{2,j}$ goes to zero and infinity,
respectively. Therefore, the ASER for the i.i.d case, given in (\ref%
{aserfinal}), can be expanded by the least power of $1/a_{2}$. Thus, from (%
\ref{ci})-(\ref{b1}), it is evident that $c_{q_{i}}^{(i)}$ is inversely
proportional to $a_{2}^{q_{i}}.$ Therefore, only the first term of the
infinite summation is considered, i.e., $q_{1}=q_{2}=l=0.$ Hence, by
plugging (\ref{a1}) and (\ref{ci}) with $m=0$ into (\ref{aserfinal}), it
yields
\begin{align}
\overline{P}_{s}^{(\text{i.i.d,}\infty )}& \sim \frac{\rho }{\pi }\left(
\frac{4\tau }{\theta a_{2}}\right) ^{N}\sum_{k_{1}+k_{2}=N}\frac{N!}{%
k_{1}!k_{2}!}\left( \frac{4}{\theta a_{2}}\right) ^{\varkappa
_{k_{1},k_{2}}}\left( b_{0}^{(1)}\right) ^{k_{1}}  \notag \\
& \times \left( b_{0}^{(2)}\right) ^{k_{2}}\mathcal{B}\left( N+\varkappa
_{k_{1},k_{2}}+\frac{1}{2},N+\varkappa _{k_{1},k_{2}}+\frac{1}{2}\right).
\label{aserasm1}
\end{align}

In (\ref{aserasm1}), only the least power of $1/a_{2}$ is kept, i.e.$,$ $%
\underset{k_{1},k_{2}}{\min }(\varkappa _{k_{1},k_{2}})+N.$ As $%
k_{2}=N-k_{1} $, it yields that $\varkappa _{k_{1},k_{2}}=\varrho
_{k_{1}}=k_{1}\left( a_{5}-a_{6}\right) +Na_{6},$ which its monotony depends
on the sign of $a_{5}-a_{6}$. As a result, its minimum corresponds to $%
k_{1}=0$ (i.e., $k_{2}=N$) if $a_{5}>a_{6},$ or $k_{1}=N$ (i.e., $k_{2}=0$)
otherwise. Consequently, it yields $\underset{k_{1}}{\min }\left( \varrho
_{k_{1}}\right) =N\min \left( a_{5},a_{6}\right) $. Lastly, by keeping only
the two aforementioned values of the pair $\left( k_{1},k_{2}\right) $ on
the multinomial summation in (\ref{aserasm1}), we obtain (\ref{aserasm}).

Analogously, and by taking only the first terms $b_{0,i}^{(k_{i})}$ $\left(
\text{i.e., }q_{1}=q_{2}=...=q_{N}=l=0\right) ,$ and using (\ref{a1}), the
ASER\ in (\ref{aserfinalinid}) can be expanded as
\begin{align}
\overline{P}_{s}^{(\text{i.n.i.d,}\infty )}& \sim \frac{\rho }{\pi }\sum
_{\substack{ k_{1}+k_{2}+...+k_{N}\geq N  \\ k_{j}=1,2}}\prod%
\limits_{j=1}^{N}\left( \tau ^{(j)}a_{2,j}^{-a_{k_{j}+4,j}-1}\right)  \notag
\\
& \times \left( \frac{4}{\theta }\right) ^{\upsilon +N} \mathcal{B}\left(
\upsilon +N+\frac{1}{2},\upsilon +N+\frac{1}{2}\right)
\prod\limits_{j=1}^{N}b_{0,j}^{(k_{j})}.  \label{aserasm1inid}
\end{align}

Thus, only the terms $a_{2,j}$ with least powers are kept. Therefore, we
define the indices $g_{j}$ satisfying $a_{g_{j}+4,j}=\min \left(
a_{5,j},a_{6,j}\right) ,$ which yields the least powers of $a_{2,j}.$
Consequently, by substituting $k_{j}$ by $g_{j}$ and $\upsilon $ by $\zeta
=\sum\limits_{j=1}^{N}a_{g_{j}+4,j}$, (\ref{aserasminid}) is achieved.
\end{IEEEproof}
\end{corollary}

{%
{\begin{remark}
From \cite[Eq. (20)]{ansarimalaga}, one can see that $\mu _{i}$ is a
function of $\alpha ,$ $\beta $, and $\xi ^{2},$ which yields that $\varphi
_{i}$ depends on the same parameters. Consequently, it yields from (\ref{a5}), (\ref{a6}), (\ref{li}), (\ref{kappa}), and (\ref{eta}) that $a_{5}$ and $a_{6}$ depend also on $\alpha ,$ $\beta $, and $\xi ^{2}.$ As a result, it
yields from (\ref{gd}) that the system's achievable diversity order for the
i.i.d case depends only on the number of branches $N$ and the turbulence and
{{PEI}} severity parameters (i.e., $\alpha ,$ $\beta $, and $\xi ^{2})
$.
\end{remark}}}

\section{Numerical Results}

In this section, some representative numerical examples are depicted in
order to highlight the effects of the key system parameters on the derived
PDF, MGFs, and ASER for both i.i.d and i.n.i.d cases. To this end, we set
the parameters $\alpha ^{(j)}=2.296$ (Except Fig. 6), $\beta ^{(j)}=2$
(Except Fig. 6), $\xi ^{(j)}=2.553$ (except Figs. 3, 5, and 6), $\Omega
=1.3265,\epsilon =0.596,d_{0}=0.1079,$ and $N=2$ (except Figs. 1-4 and Fig.
6). For simplicity, we define the sets $\overline{\alpha }=\{\alpha
^{(j)}\}_{1\leq j\leq N},$ $\overline{\beta }=\{\beta ^{(j)}\}_{1\leq j\leq
N},$ $\overline{\xi }=\{\xi ^{(j)}\}_{1\leq j\leq N}.$The simulation was
performed by generating $3\times 10^{6}$ MRV-distributed random samples per
each average SNR value.

\begin{table*}[th]
\vspace{-.675em} \hspace{-2.7em}
\begin{tabular}{p{7cm}p{7cm}p{7cm}}
\hspace*{-.14cm}\includegraphics[scale=0.50]{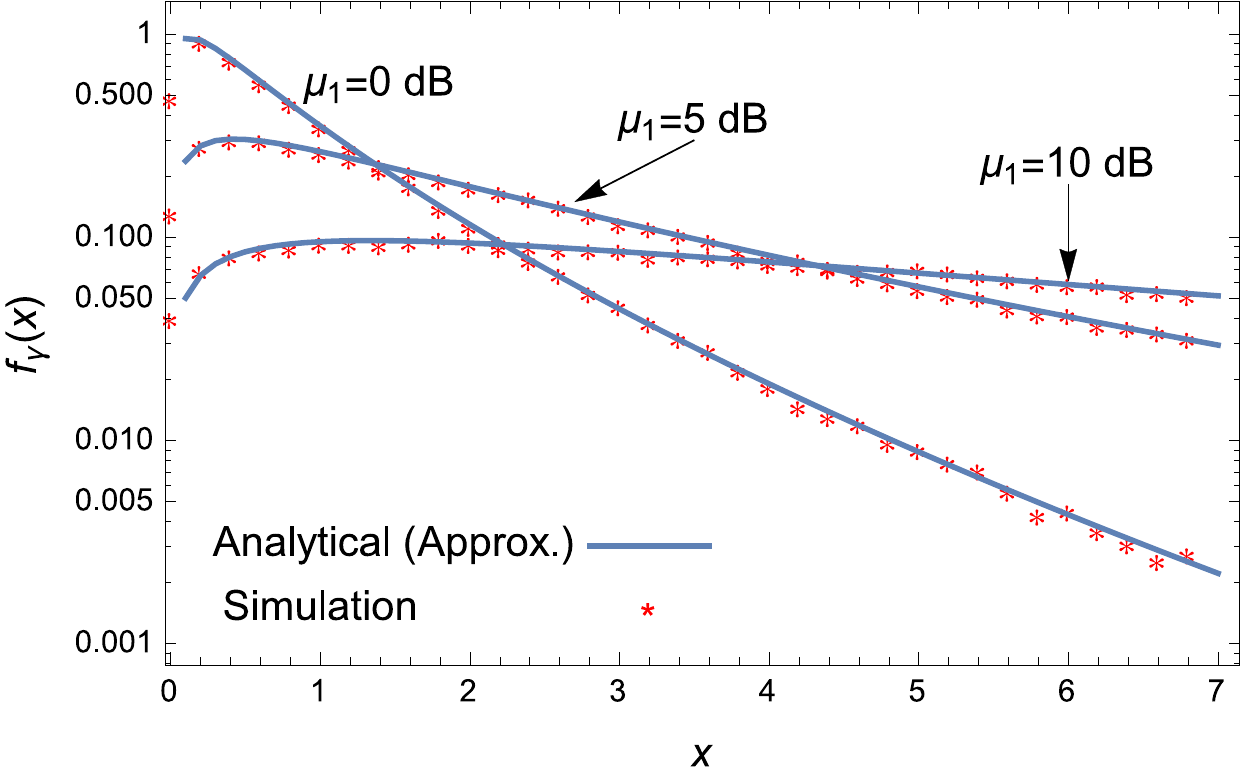} & \hspace*{-1cm}%
\includegraphics[scale=0.50]{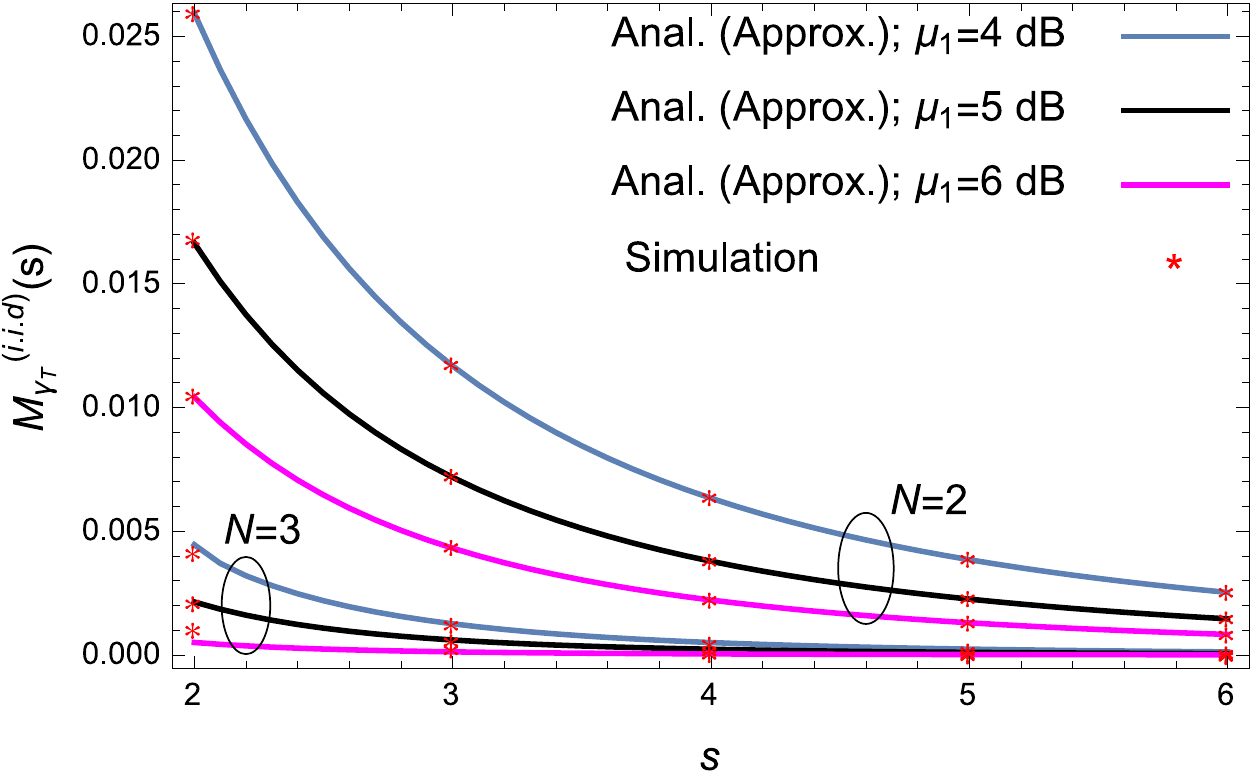} & \hspace*{-1.82cm}%
\includegraphics[scale=0.50]{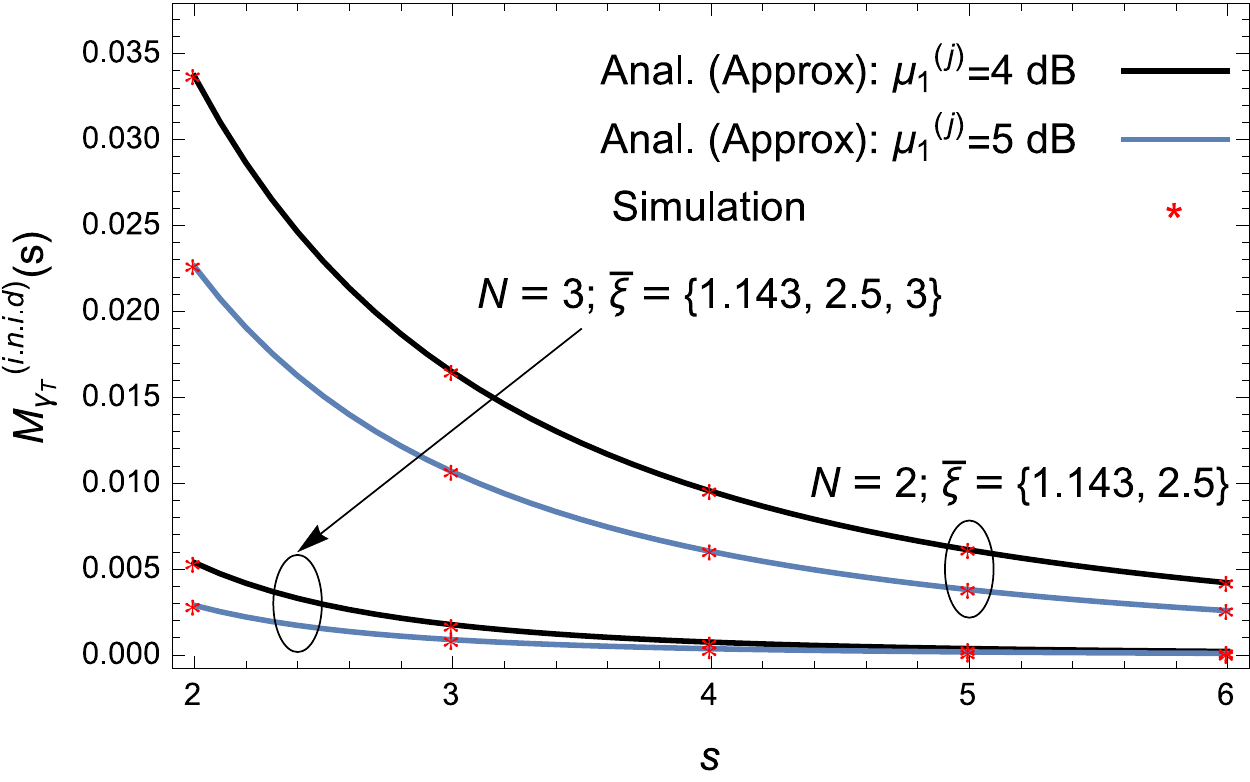} \\
{\hspace*{.3cm}{\footnotesize {{{Fig. 1. Approximate PDF\ of MRV.}%
}}}} & \hspace*{-.77cm}{\footnotesize {{{Fig. 2. MGF\ of the sum\
of i.i.d MRVs.}}}} & \hspace*{-1.68cm}{\footnotesize {{{Fig. 3.
MGF\ of the sum\ of i.n.i.d MRVs.}}}} \\
\vspace*{1cm}\hspace*{-.11cm}\includegraphics[scale=0.50]{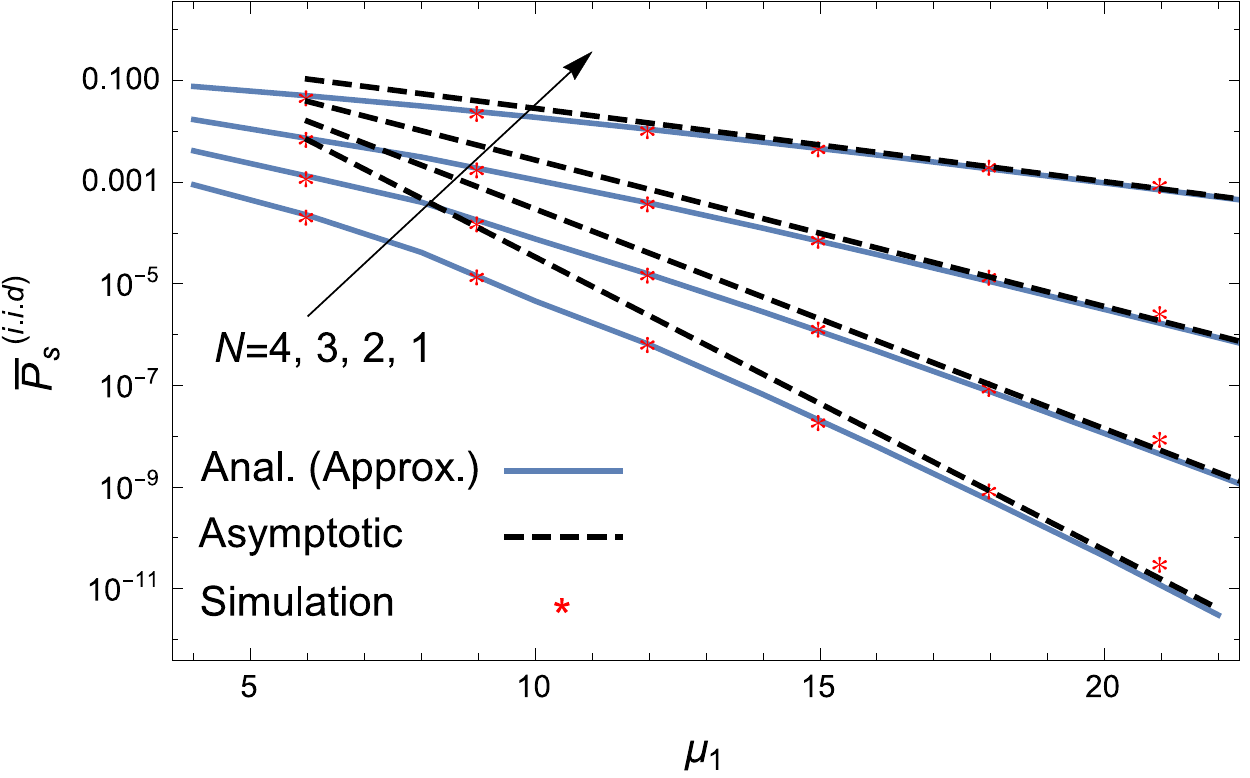} &
\vspace*{1cm}\hspace*{-.95cm}\includegraphics[scale=0.50]{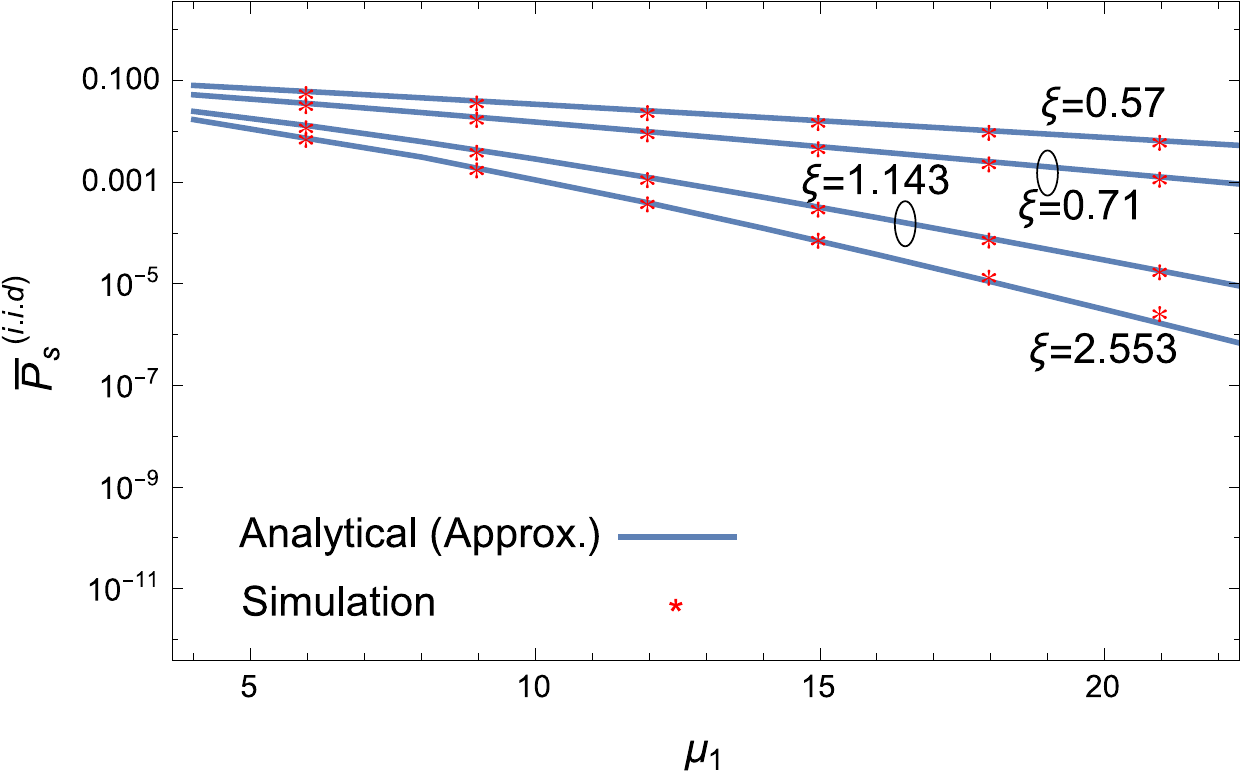} &
\vspace*{1cm}\hspace*{-1.78cm}\includegraphics[scale=0.50]{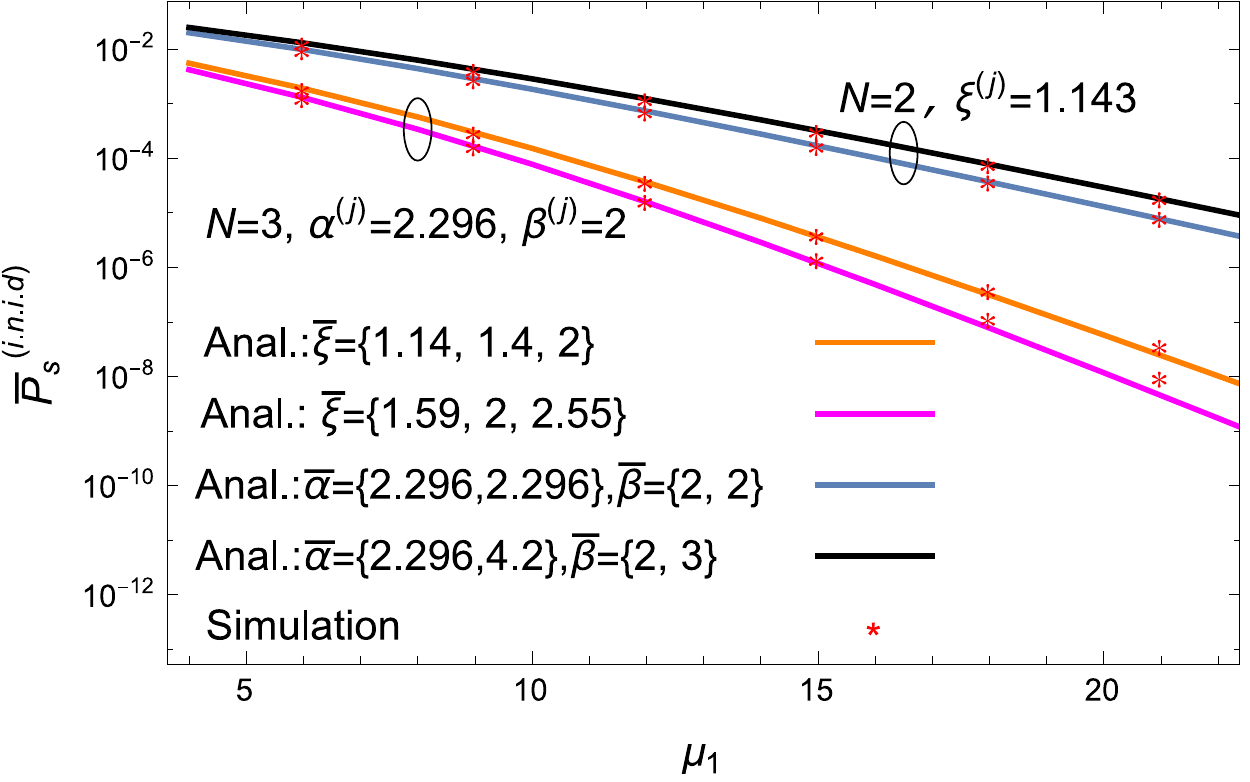} \\
\hspace*{.4cm}{\footnotesize {{{Fig. 4. ASER\ vs $\mu_1$ for the
i.i.d case.}}}} & \hspace*{-.9cm} {\footnotesize {{{Fig. 5. ASER
vs $\mu_1$ for the i.i.d case with $N=2$.}}}} & \hspace*{-1.63cm}%
{\footnotesize {{{Fig. 6. ASER\ vs $\mu_1$ for the i.n.i.d case.}}%
}}%
\end{tabular}%
\end{table*}

Fig. 1 depicts the exact simulated and approximate PDF, computed using (\ref%
{malagapdf}) and (\ref{pdfapprox}), respectively. One can ascertain that the
two curves match tightly over the entire range of $x$ and for several $\mu
_{1}$ values, showing the accuracy of (\ref{pdfapprox}). Importantly, the
curves shift with $\mu _{1},$ and that the smaller $\mu _{1}$ is, the more
condensed are the random samples near to $0$, leading to a higher peak of
the PDF.

In Figs. 2 and 3, the MGF\ of the sum of $N$ MRVs, given in (\ref{mgftotal}%
)-(\ref{mgftotalinid}), is plotted alongside its Monte Carlo simulation
counterpart for both i.i.d and i.n.i.d cases. Again, it is clear that the
two functions are matching for several values of $\mu _{1}^{(j)}$ and $N$. {%
{One can note evidently that the higher $s$ is, the lower is the
MGF}}. Additionally, one can see that, for a fixed value of $s$, the higher
the average electrical SNR {{$\mu_{1}$}}, $\mu _{1}^{(j)}$ and $N$
are, the lower the MGF value. Indeed, the greater {%
{these three
parameters are}}, the higher the total average SNR $\overline{\gamma }_{T}.$
Therefore, {{the more spread is the range of $\gamma_T$}}.
Consequently, the MGF, defined as $M_{\gamma _{T}}(s)=\mathbb{E}\left[
e^{-s\gamma _{T}}\right] $ , decreases significantly. {%
{Furthermore, it is seen also that for higher $N$ values (i.e.,
$N=3$), the impact of $\mu_1$ and $\mu^{(j)}_1$ is less significant on the
MGF, particularly at higher values of $s$. This can be interpreted as the product $s\gamma_{T}$  is sufficiently higher to vanish the exponential term.}%
}

Fig. 4 highlights the ASER\ versus$\ \mu _{1}$ of an $N$-branch MRC\ WCS,
subject to i.i.d M\'{a}laga-$\mathcal{M}$ fading with {{PEI}}, as
given in (\ref{aserfinal}), alongside its asymptotic result in (\ref{aserasm}%
). The ASER\ is plotted for BPSK\ modulation for various $N$ values. It is
evident that the analytical curves are tightly close to the Monte\ Carlo\
simulation ones, particularly for SNR\ values below $20$ dB. Additionally,
the greater $\mu _{1}$ and $N$ are, the enhanced is the overall system's
ASER performance.

Fig. 5 depicts the ASER\ for the i.i.d fading case with $N=2$ and several
values of $\xi .$ It can be noticed that the higher the $\xi $ value, the
lesser is the {{PEI}} effect, leading to an enhancement of the
system's performance.

Fig. 6 shows the ASER\ versus$\ \mu _{1}=\mu _{1}^{(j)}$ for the i.n.i.d
scenario, given in (\ref{aserfinalinid}), for various $\alpha ^{(j)},$ $%
\beta ^{(j)}$, and $\xi ^{(j)}$ values. One can notice that the higher $%
\alpha ^{(j)}$ and $\beta ^{(j)}$ per branch, the overall system's ASER
improves. Indeed, the higher these two parameters are, the lesser the
turbulence effect. Again, the smaller $\xi ^{(j)}$ per branch, the worse is
the system's performance.

\section{Conclusion}

In this work, an accurate approximation {{for the statistics
of}} the sum\ of MRVs was proposed. In particular, we proposed an
approximation for the PDF\ of MRV, using the moments-based approach, by
considering the first six moments, making it highly accurate. Next, the MGF
of the sum of MRVs was retrieved, based on which, the respective ASER\ for
MRC\ combining scheme was derived in approximate and asymptotic forms, for
both i.i.d and i.n.i.d fading scenarios. The assessed analysis showed that
the proposed approximation yields a very close result to the exact
simulation ones, for several values of the system's parameters.
Additionally, the system's ASER performance\ is clearly impacted by the
turbulence and {{PEI}} parameters. Lastly, the results showed that
the achievable diversity order increases by increasing the number of the
receiver's branches.

{{
\section*{Appendix A: Proof of Proposition 1}

The moment-based approximation approach is opted in this work to retrieve an
approximate RV$\ \widehat{\gamma }$ for $\gamma $. It is worth noting that
the greater the number of satisfied equations $\mathbb{E}\left[ \widehat{%
\gamma }^{i}\right] =\mathbb{E}\left[ \gamma ^{i}\right] $ $(0\leq i\leq
K-1) $, the tighter is such an approximation. Owing to this fact, we choose $%
K=6$ instead of $K=5$ opted in \cite{wcl}. In a similar manner to the
last-mentioned work, $\widehat{\gamma }$ is also considered in this work an $%
H$-distribution with $K$ parameters $\left\{ a_{i}\right\} _{1\leq i\leq K}$%
, among them $a_{1}$ and $a_{2}$ are the scale factors of the PDF and $x$,
respectively$,$ as shown in (\ref{pdfapprox}). Moreover, the remaining ones
are split between the integrand terms of the corresponding Mellin-Barnes
integral, so as to reduce, relying on \cite[Eq. (8.331.1)]{integrals}, the
complexity of the following system of linear equations, obtained with the
aid of the Mellin transform \cite[Eq. (2.9)]{mathai}%
\begin{equation}
\mu _{i}=a_{1}a_{2}^{i+1}\frac{\Gamma \left( a_{5}+i+1\right) \Gamma \left(
a_{6}+i+1\right) }{\Gamma \left( a_{3}+i+1\right) \Gamma \left(
a_{4}+i+1\right) };0\leq i\leq K-1.
\end{equation}

To this end, by considering that the moment of order $i=0$ equals unity
(i.e. $\mu _{0}=1),$ (\ref{a1}) can be reached. Additionally, using \cite[%
Eq. (8.331.1)]{integrals}, we get
\begin{equation}
\mathcal{L}_{i}-2\mathcal{L}_{i-1}+\mathcal{L}_{i-2}=2a_{2};i\geq 3.
\label{combin}
\end{equation}

Thus, (\ref{a2}) can be readily obtained by setting $i=4$ in (\ref{combin}).
On the other hand, by putting $i$ equal to $3$ and $4$ into (\ref{combin}), {%
{two equations are obtained. To this end, by}} making equality
between the left hand-sides of the two obtained equations, alongside some
algebraic manipulations, (\ref{a3}) is obtained. Analogously, by
substituting $i=4$ and $5$ in (\ref{combin}), {%
{and making
equality between the left hand-sides of the two obtained equations,}} one
obtains (\ref{combinp}), as shown at the bottom of the page.
\begin{figure*}[b]
{\normalsize \setcounter{mytempeqncnt}{\value{equation}} %
\setcounter{equation}{44} }
\par
{{\begin{equation}
a_{3}a_{4}\left[ \varphi _{5}-3\varphi _{4}+3\varphi _{3}-\varphi _{2}\right]
+\left( a_{3}+a_{4}\right) \left[ 5\varphi _{5}-12\varphi _{4}+9\varphi
_{3}-2\varphi _{2}\right] +25\varphi _{5}-48\varphi _{4}+27\varphi
_{3}-4\varphi _{2}=0.  \label{combinp}
\end{equation}}}
\par
{\normalsize \hrulefill \vspace*{4pt} }
\end{figure*}

Now, by plugging (\ref{a3}) into (\ref{combinp}) and performing some
algebraic operations, it yields (\ref{a4}). Moreover, we have the two
following identities
\begin{equation}
\left\{
\begin{array}{l}
a_{5}+a_{6}+2i+1=\frac{\mathcal{L}_{i+1}-\mathcal{L}_{i}}{a_{2}} \\
\left( a_{5}+1\right) \left( a_{6}+1\right) =\frac{\mathcal{L}_{1}}{a_{2}}%
\end{array}%
\right. .  \label{combin2}
\end{equation}

By setting $i=1$ in (\ref{combin2}), $a_{5}$ and $a_{6}$ are attained as
shown in (\ref{a5}) and (\ref{a6}), respectively. }}





%
\bibliographystyle{IEEEtran}
\bibliography{sigproc}


\end{document}